\documentclass[letters,fleqn,usenatbib]{mnras}

\usepackage{newtxtext,newtxmath}

\usepackage[T1]{fontenc}

\DeclareRobustCommand{\VAN}[3]{#2}
\let\VANthebibliography\thebibliography
\def\thebibliography{\DeclareRobustCommand{\VAN}[3]{##3}\VANthebibliography}

\usepackage{graphicx}	
\usepackage{amsmath}	

\usepackage{orcidlink}

\title[$\Delta$$m$\textsubscript{10}($X$)]{Shortening the Baseline: Characterising Type Ia Supernovae Using $\Delta m_{10}(X)$}

\author[J. M. Velasco et al.]{Joel M. Velasco,$^{1}$\thanks{E-mail: arijvela@ljmu.ac.uk} \orcidlink{0009-0000-5092-4090}
Paolo A. Mazzali,$^{1,2}$ \orcidlink{0000-0001-6876-8284}
Christopher Ashall,$^{3}$ \orcidlink{0000-0002-5221-7557}
Christopher R. Burns,$^{4}$ \orcidlink{0000-0003-4625-6629}
\newauthor
Mark M. Phillips,$^{5}$ \orcidlink{0000-0003-2734-0796}
Lluis Galbany,$^{6,7}$ \orcidlink{0000-0002-1296-6887} 
Nidia Morrell,$^{5}$ \orcidlink{0000-0003-2535-3091}
Maximilian D. Stritzinger,$^{8}$ \orcidlink{0000-0002-5571-1833}
\newauthor
Peter Hoeflich,$^{9}$ \orcidlink{0000-0002-4338-6586}
Eric Y. Hsiao,$^{9}$ \orcidlink{0000-0003-1039-2928}
Syed A. Uddin,$^{10,11}$ \orcidlink{0000-0002-9413-4186}
\\
$^{1}$Astrophysics Research Institute, Liverpool John Moores University, IC2, Liverpool Science Park, 146 Brownlow Hill, Liverpool, L3 5RF, UK\\
$^{2}$Max-Planck-Institut für Astrophysik, Karl-Schwarzschild-Str. 1, 85748 Garching, Germany\\
$^{3}$Institute for Astronomy, University of Hawaii, 2680 Woodlawn Dr, Honolulu, HI 96822, USA\\
$^{4}$Observatories of the Carnegie Institution for Science, 813 Santa Barbara St., Pasadena, CA 91101, USA\\
$^{5}$Carnegie Observatories, Las Campanas Observatory, Colina El Pino, Casilla 601, Chile\\
$^{6}$Institute of Space Sciences (ICE-CSIC), Campus UAB, Carrer de Can Magrans, s/n, E-08193 Barcelona, Spain\\
$^{7}$Institut d'Estudis Espacials de Catalunya (IEEC), 08860 Castelldefels (Barcelona), Spain\\
$^{8}$Department of Physics and Astronomy, Aarhus University, Ny Munkegade 120, DK-8000 Aarhus C, Denmark\\
$^{9}$Department of Physics, Florida State University, 77 Chieftan Way, Tallahassee, FL 32306, USA\\
$^{10}$Center for Astronomy, Space Science and Astrophysics, Independent University, Bashundhara R/A, Dhaka 1229, Bangladesh\\
$^{11}$American Public University System, 111 W. Congress St., Charles Town, WV 25414, USA\\
}
\date{Accepted XXX. Received YYY; in original form ZZZ}

\pubyear{\the\year{}}

\begin{document}
\label{firstpage}
\pagerange{\pageref{firstpage}--\pageref{lastpage}}
\maketitle

\begin{abstract}
An explorative study of Type Ia Supernovae (SNe Ia) light curves is presented. $\Delta m_t(X)$ decline-rate metrics are used to evaluate alternative characterisations of light curve diversity and the Luminosity-Width Relation (LWR). Using a sample of 362 SNe Ia from the Carnegie Supernova Project (CSP-I and CSP-II), Gaussian Process regression is used to fit $B$- and $g$-band light curves and measure decline-rates across multiple post-maximum timescales. The LWR is calibrated using a clean subsample of SNe Ia with host-independent distance moduli and Fitzpatrick (1999) reddening corrections. Decline-rates $\Delta m_{10}(B)$ and $\Delta m_{10}(g)$ are found to capture the LWR effectively, while mitigating the severe degeneracy that affects longer timescales for fast-declining ($\Delta m_{15}(B) \ge 1.6\,\text{mag}$) events. Continuous, piecewise linear models are presented that seamlessly anchor the transition between normal and fast-declining regimes. 
The decline-rate $\Delta m_{10}(X)$ closely approximates the results of the more complex colour-stretch parameters, requiring less post-maximum temporal coverage than 
$\Delta m_{15}(B)$ or $s_{BV}$, and offers an even greater sensitivity for slow decliners. 
$\Delta m_{10}(B)$ provides a precise and practical tool for large-scale photometric surveys and real-time observer classification.
\end{abstract}

\begin{keywords}
supernovae: general -- techniques: photometric -- methods: data analysis -- cosmological parameters
\end{keywords}



\section{Introduction}
\label{sec:intro} 

Type Ia Supernovae (SNe Ia) are luminous stellar explosions which can be observed out to redshifts corresponding to a time when the Universe was half its present age \citep{1998Natur.391...51P}. SNe Ia have been utilised as indicators of extragalactic distance thanks to the discovery that the rate of evolution of their light curves correlates with their intrinsic luminosity \citep{1977SvA....21..675P, 1993ApJ...413L.105P}. They were instrumental in the discovery of the accelerated expansion of the Universe \citep{Riess_1998,1999ApJ...517..565P}. Despite this, much is yet to be learned about their physical properties, and even the nature of their progenitors (beyond the presence of at least one exploding carbon-oxygen white dwarf) is a matter of debate \citep{hoyle1960ApJ...132..565H,Wang_2012, Livio_2018}.

To correct for an apparent Luminosity-Width Relation (LWR), \citet{1993ApJ...413L.105P} devised a measure of the decline-rate, $\Delta m_{15}(B)$. Defined as the $B$-band magnitude difference from peak to 15 days later, this parameter has been crucial for characterising the intrinsic luminosity of SNe Ia and enabling their use as standard candles. Historically, the 15-day baseline was a practical compromise dictated by the photometric cadence of the Calan/Tololo survey, where shorter timescales weren't feasible \citep{CalanHamuyPhillips}. When applied alongside a colour-luminosity relation \citep{1999AJ..riess..117..707R,1998A&A...331..815T}, SNe Ia can be standardised, reducing peak luminosity scatter to $\sim 0.15$ mag \citep{Phillips1999AJ....118.1766P,2007ApJ...659..122J}. However, while this timebase captures the variance of most SNe Ia, it struggles to distinguish between the fastest-declining events. 

$\Delta m_{15}(B)$ exhibits a prominent breakdown at the faint, rapidly declining end of the luminosity distribution ($\Delta m_{15}(B) \ge 1.6\,\text{mag}$; \citealt{2008MNRAS.385...75T, folatelli2010}). These fast-declining, 1986G and 1991bg-like SNe Ia synthesise lower $^{56}\mathrm{Ni}$ masses \citep{Mazzali07}, leading to cooler ejecta, earlier Fe-group recombination, and severe line-blanketing by Ti\,\textsc{ii} that rapidly depresses post-maximum $B$-band flux \citep{10.1093/mnras/284.1.151, 2026Alaa}. At $\Delta m_{15}(B) \ge 1.6$,mag, $\Delta m_{15}(B)$ becomes highly degenerate, with SNe Ia of comparable decline-rates spanning more than two magnitudes in absolute luminosity \citep{Burns_2014, Phillips2017}. The increasing difficulty of tracking luminosity over longer timescales suggests there is some impact felt, even in the $B$ band, from the onset of the secondary maximum (Velasco et al., in prep.). While the secondary maximum is conventionally associated with near-infrared wavelengths, our analysis reveals a suppressed decline extending into the optical bands. For the dimmest, fastest-declining SNe Ia, the colour turnover driven by early Fe-group recombination occurs as early as $\sim 13$ days post-maximum. Because this transition takes place before the traditional 15-day measurement window for fast-decliners, $\Delta m_{15}(X)$ becomes fundamentally degenerate.

To address these degeneracies, other luminosity proxies have been developed, for instance the colour-stretch parameter, $s_{BV}$ \citep{Burns_2014,Burns_2018}. This parameter successfully characterises SNe Ia across the full luminosity range by utilising the $(B-V)$ colour curve. To measure $s_{BV}$ directly from light curves requires high-cadence, multi-band photometry extending 50 days past the $B$ band maximum.

Given this state of affairs, it was natural to ask the question whether using a shorter timescale to measure the decline-rate would eliminate contamination from the secondary maximum and remove the degeneracy at the rapidly declining end, while retaining sensitivity to the absolute magnitude for normal SNe Ia.

\citet{folatelli2010} experimented with modifying $\Delta$$m$\textsubscript{15}($B$), altering both timescale and photometric band. However, the dataset used was limited, containing only two fast declining SNe Ia. Similarly, \citet{Hoef2017}  studied decline-rates measured at a variable times after maximum, stretching light curves prior to measuring their decline. Here, we experiment with the decline-rate $\Delta$$m$\textsubscript{15}($B$), modifying the 15 day period, whilst also measuring declines in the $g$ band. Employing modern data that include a greater number of SNe Ia from across the luminosity range, we aim to find alternative timescales which more accurately encapsulate the variety in SNe Ia light curves, without needing the larger temporal baseline of $s_{BV}$ or fitting with light curve templates.

\section{Data}
\label{sec:data}

The data used in this letter are from the Carnegie Supernova Project (CSP), with this sample spanning both CSP-I \citep{Krisciunas_2017} and CSP-II surveys \citep{Phillips_2019}. The CSP-I data include the photometry of 133 SNe Ia observed between 2004 and 2009. A further 229 SNe Ia are from CSP-II observations, which took place from 2011 to 2015 \citep{Phillips_2019}. For CSP-I and II, optical imaging was obtained with the Las Campanas Observatory Swope telescope, with each SN observed every 2-3 days up to 2-3 weeks past maximum light. Combining the data from CSP-I and CSP-II into one larger dataset, a total of 362 SNe Ia are used for analysis here. Supplementary parameters for the host galaxies, specifically the colour excess ($E(B-V)_{\text{host}}$) and the total-to-selective extinction coefficient ($R_{V,\text{host}}$), were derived from multi-wavelength template fits using \textsc{SNooPy} \citep{Burns2011AJ....141...19B, Burns_2014,Uddin_2024}. Spectroscopic subclassifications were adopted from the literature. SNe Ia-91T were defined in \citet{Phillips_2022}, and 03fg-like SNe Ia (SNe Ia-03fg) are defined from \citet{ashall21}.

In order to calibrate the absolute magnitudes of SNe Ia in the set, a calibration subsample of 23 CSP Ia events with reliable, redshift independent distance moduli was assembled. Cepheid distances were given priority, followed by tip of the red giant branch (TRGB), surface brightness fluctuations (SBF) and Tully-Fisher (TF) distances. Distance moduli were found using the Extragalactic Distance Database \citep{EDD}, with data taken from the SH0ES \citep{SH0ES}, Cosmicflows-4 \citep{cosmicflows4} and HST IR SBF catalogues \citep{SBF.Blakeslee,SBF.Jensen}. For the remainder of the dataset (the `flow' sample), independent absolute magnitudes were computed from their cosmic microwave background (CMB) centric redshifts and the Hubble Law, assuming $H_0 = 73.04\,\text{km}\,\text{s}^{-1}\,\text{Mpc}^{-1}$ \citep{SH0ES}.

\section{Methodology}
\label{sec:methods} 

To ensure reliable decline-rate measurements, we required all SNe to possess photometric data prior to their respective band maximum. Because the peak must be well-sampled to measure accurate decline-rates, we defined the epoch of maximum light using the peak of the fitted light curve rather than a single discrete data point, maximising the number of SNe eligible for analysis.

\begin{figure}
     \centering
         \centering
         \includegraphics[width=0.48\textwidth]{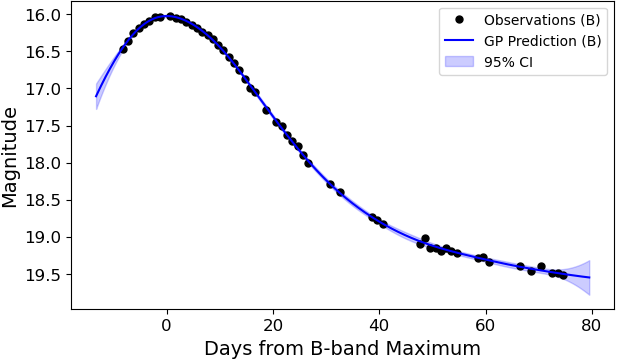}
        \caption{An example of a fitted light curve for SN2005M's $B$ band data. The black dots show the photometric data points and the blue line shows the GP fitted light curve, alongside a 95\% confidence interval around the fit.}
        \label{fig:Example Light curves}
\end{figure}

To preserve the true morphological heterogeneity of the SNe Ia, we refrained from using templates to measure the decline-rates. By directly fitting the photometric data points alone, the intrinsic nature of the SN light curve shape is preserved. A Gaussian Process (GP) regressor, utilising a Matérn kernel, implemented via the \textsc{scikit-learn} and \textsc{scipy} Python modules, was used to predict the shape of the light curves through probabilistic interpolation. All light curves were corrected for cosmological time dilation; K-corrections were deemed unnecessary given the low redshift of our sample ($z_{max}<0.14$), with around $75\%$ having $z<0.05$. An example of a high-quality $B$-band light curve accurately fit by the GP model, alongside its 95\% confidence interval, is shown in Fig.~\ref{fig:Example Light curves}.

\begin{figure*}
\centering
\includegraphics[width=0.72\textwidth]{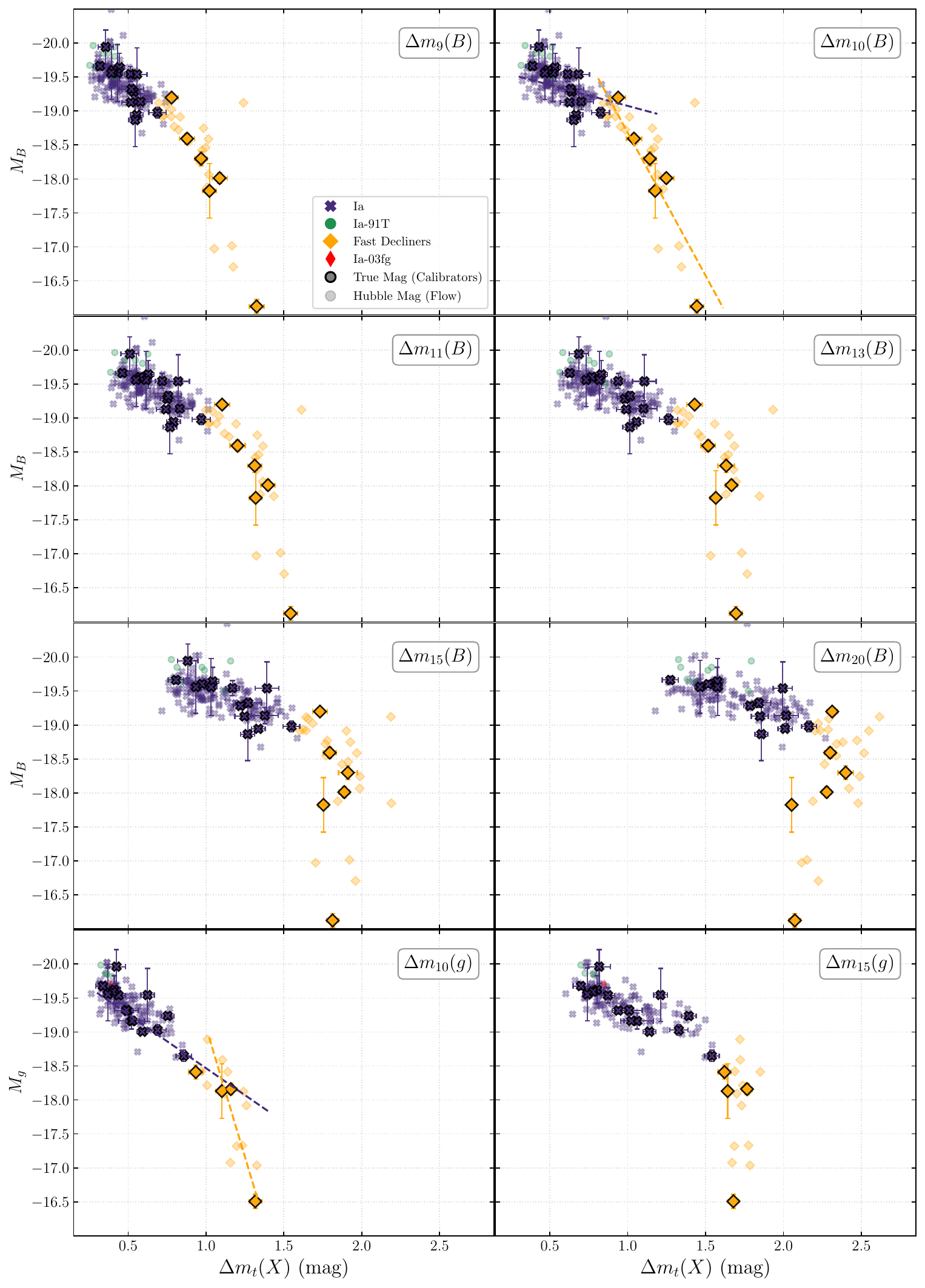} 
\caption{Absolute magnitude vs $\Delta m_t(X)$ for a variety of timescales in both the $B$ and $g$ bands. True empirical calibrator SNe are highlighted with bold black outlines, while the larger `flow' sample is plotted as a faded background distribution, using Hubble-derived distances. Error bars for the flow sample are omitted for visual clarity. The derived continuous piecewise linear models (dashed lines) are overlaid on the $\Delta m_{10}(X)$ panels.}
\label{fig:phil grid}
\end{figure*}

From the GP fits, the time of maximum brightness in the $B$ and $g$ bands was extracted, allowing various decline-rates to be calculated. We determine uncertainties on $\Delta m_t(X)$ using a Monte Carlo approach based on the GP posterior. For each SN, we draw 100 random realisations from the GP fit, independently find the peak time ($t_{X,\text{max}}$) for each iteration, and compute the resulting decline-rates. Taking the standard deviation across all 100 realisations maps the $t_{\text{max}}$ covariance into the uncertainty.

Following the naming convention of \citet{folatelli2010}, these are referred to as $\Delta m_t(X)$, where $t$ denotes the number of days since maximum $X$-band luminosity. $\Delta m_t(X)$ was computed for values of $t$ from 8 to 20 days. Following the application of our selection criteria, we obtained $B$-band measurements for 224 SNe at $t=8$ and 191 SNe at $t=20$. In the $g$-band, the yield ranged from 162 SNe at $t=8$ to 153 SNe at $t=20$.

Additionally, $(B-V)$ and $(g-r)$ colour curves were calculated to derive contemporary colour-stretch parameters. The $(B-V)$ curve was filtered for post-maximum epochs, and the time of the reddest point was extracted and divided by 30 to obtain $s_{BV}$ \citep{Burns_2014, Burns_2018}. We measured $s_{BV}$ directly for 81 SNe. For the $g$-band, the equivalent parameter $s_{gr}$ \citep{ashall21} (the time of $(g-r)$ maximum relative to $g$-band maximum, normalised by 30 days) was computed for 76 SNe.

To calibrate the absolute magnitudes, we utilised the independent subsample of 23 calibrator SNe introduced in Sec.~\ref{sec:data}. Peak absolute magnitudes For these SNe were derived directly from the GP fits:
\begin{align*}
M_X = m_{X,\text{max}} - \mu - A_{X,\text{MW}} - A_{X,\text{host}},
\end{align*}
where $m_{X,\text{max}}$ is the unreddened peak apparent magnitude. Total Galactic and host extinctions ($A_X$) were computed using the \citet{F99} extinction law, assuming $R_V = 3.1$ for the Milky Way, alongside the host galaxy parameters ($E(B-V)_{\text{host}}$ and $R_{V,\text{host}}$) noted in Sec.~\ref{sec:data}. Milky Way colour excesses, $E(B-V)_{\text{MW}}$, were obtained from \citet{Schlafly_2011} dust maps. 

Using this calibrator sample, we constructed empirical LWRs for both the $B$ and $g$ bands. To accurately model the distinct photometric behaviour of fast-decliners without introducing discontinuities, we adopted an anchored, continuous piecewise linear fit. The transition into the fast-declining regime was empirically defined by the threshold $\Delta m_{15}(B) \ge 1.6$\,mag \citep{2008MNRAS.385...75T, folatelli2010, Ashall_2016}. The slopes and transition points of these piecewise models were determined exclusively using the 23 calibrator SNe.

\section{Results}
\label{sec:results} 

From these data, $\Delta$$m$\textsubscript{t}($X$) decline-rates were plotted in order to infer their correlation with luminosity. This was in order to test for a reproduction of the LWR \citep{1993ApJ...413L.105P}, whilst also analysing how the distribution changes for different values of $t$. Plots depicting absolute magnitude against $\Delta$$m$\textsubscript{t}($X$) can be seen in Fig.~\ref{fig:phil grid}.

The $\Delta$$m$\textsubscript{15}($B$) plot appears as expected; for normal SNe Ia ($\Delta m_{15}(B) < 1.6$\,mag), the decline-rate correlates reliably with peak luminosity. However, as the decline-rate crosses the $\Delta m_{15}(B) \approx 1.6$\,mag threshold, this relationship breaks down entirely. The degeneracy caused by the fast-decliners means that assuming a luminosity from $\Delta$$m$\textsubscript{15}($B$) alone for SNe with $\Delta$$m$\textsubscript{15}($B$)$>1.6$\,mag could lead to discrepancies greater than 2 mag. Such an uncertainty highlights the motivation and opportunity for another metric, one which maintains resolving power over a greater range of luminosities.

For the lowest values of $t$ ($\le11$), a larger value of $\Delta$$m$\textsubscript{$t$}($B$) corresponds to decreasing luminosity (although a degree of degeneracy is still present). The two-part piecewise model (whose tracks can be seen plotted, crossing over in the transitional region) resolves the degeneracy seen in the $\Delta$$m$\textsubscript{15}($B$) panel. The two distinct tracks, anchored by the calibrator SNe, are continuous, with the flow sample scattering naturally around them.

Intermediate values of $t$ (i.e. $t$ = 13 or 15) show a larger value of $\Delta$$m$\textsubscript{$t$}($B$) correlating with decreasing luminosity until the knee is met. At this point the fast decliners show no clear trend, instead displaying a degeneracy wherein differences of $\Delta$$m$\textsubscript{$t$}($B$) show no discernable indication of luminosity. At its most extreme, there is over a 2.5\,mag difference between points of equal $\Delta$$m$\textsubscript{$15$}($B$).

With $t=20$, the largest value tested in this letter, slow decliners show the same pattern as always (increasing $\Delta$$m$\textsubscript{$t$}($B$) correlates with decreasing magnitude). Fast decliners, on the other hand, show the inverse; an increase in $\Delta$$m$\textsubscript{$t$}($B$) shows an increase in luminosity. Comparing to $\Delta$$m$\textsubscript{15}($B$), the break shown by fast decliners in the $t=20$ plot is far more accentuated. Nevertheless, this extended timescale does offer a distinct advantage for the normal and slow-declining populations: their decline-rates become distributed over a significantly wider range of values. This flattens the slope of the relation, meaning that small measurement uncertainties in the decline-rate (e.g. $\sim 0.05$\,mag) have a minimal impact on the derived absolute magnitude.

We also tested whether these relations hold in the $g$-band, given its similarity in wavelength to the $B$ band and the fact that modern surveys often prefer its use. Results are shown in the bottom panels of Fig.~\ref{fig:phil grid}. The correlations that are obtained are consistent with their $B$-band counterparts; the fastest-declining SNe remain degenerate compared to the rest of the data for $t=15$. $\Delta$$m$\textsubscript{10}($g$) does present a steeper relation for fast-decliners than $\Delta$$m$\textsubscript{10}($B$), but despite that a severe degeneracy is still mitigated. Note also that the transitional fast-decliners seamlessly joining the slow-declining track in the $g$-band.

\begin{figure}
\centering
\includegraphics[width=0.48\textwidth]{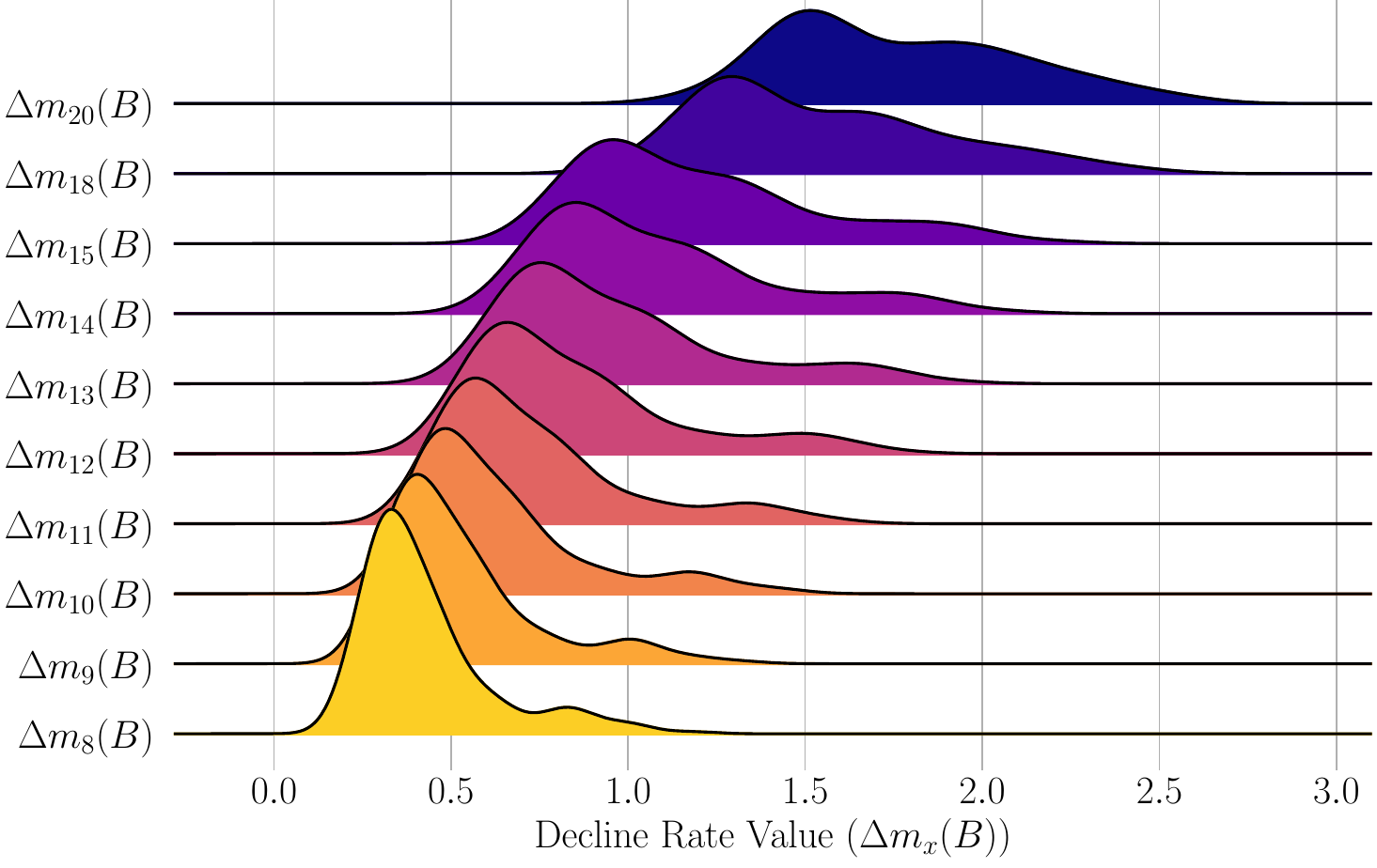} 
\caption{Ridgeline plot showing the distributions of $\Delta$$m$\textsubscript{$t$}($B$) decline-rates.}
\label{fig: ridge}
\end{figure}

Studying the $\Delta$$m$\textsubscript{$t$}($B$) distribution as a whole, Fig.~\ref{fig: ridge} shows a ridgeline plot that demonstrates how the frequency of $B$ band decline-rates changes as the timescale is altered. For the shortest timescales, it can be seen that there is one dominant population with a slight knee populated by the fastest-decliners. As the timescale increases, the dominant population starts to spread over a larger range of values, and the knee becomes more accented, with the fastest-declining SNe separating themselves even further from the larger population. At a timescale of around 11 days the spread in the dominant population evolves into signs of a second bump, potentially a bimodality in $\Delta$$m$\textsubscript{$15$}($B$), as seen in \citet{Ashall_2016}. However, given that the data used here are not a complete, volume-limited sample, such a conclusion is weak. The bump becomes more defined with time until the distribution starts to appear more like three similar, albeit slightly different populations. The dominant population (consisting of the slow-declining extreme and normal Ia's) and the extended knee, corresponding to the fast-decliners, remain clearly defined, while an additional population sitting between the others also becomes more defined. 

The SNe populating this region are comprised of faster declining normal SNe Ia, and their behaviour on this plot is fascinating when comparing to their positions on LWR plots for instance (Fig.~\ref{fig:phil grid}). Where those plots suggest that SNe Ia are separated into groups of 91bg-like and non-91bg-like, the distribution of SNe at the largest timescales on this plot ($t=20$ in particular) is in opposition to this conclusion. Instead there appears to be an almost continuous distribution where the fastest and slowest-decliners are blended into one another by the fast declining normal Ia's populating the aforementioned bump. One could either look at this as evidence of a continuous class of SN, or instead see it as an example of SNe Ia discontinuity where two (or possibly three) populations are being treated as one.

Though $\Delta$$m$\textsubscript{$t$}($B$) for the full range of $t$ is shown in this letter, isolating the most successful decline-rate allows for a more focused comparison against the already established $\Delta$$m$\textsubscript{$15$}($B$). $\Delta$$m$\textsubscript{$t$}($B$) where $t\le11$ are the clear candidates, resulting from the absence of a defined break section in these plots in Fig.~\ref{fig:phil grid} compared to larger values of $t$.

\begin{figure*}
\centering
\includegraphics[width=0.86\textwidth]{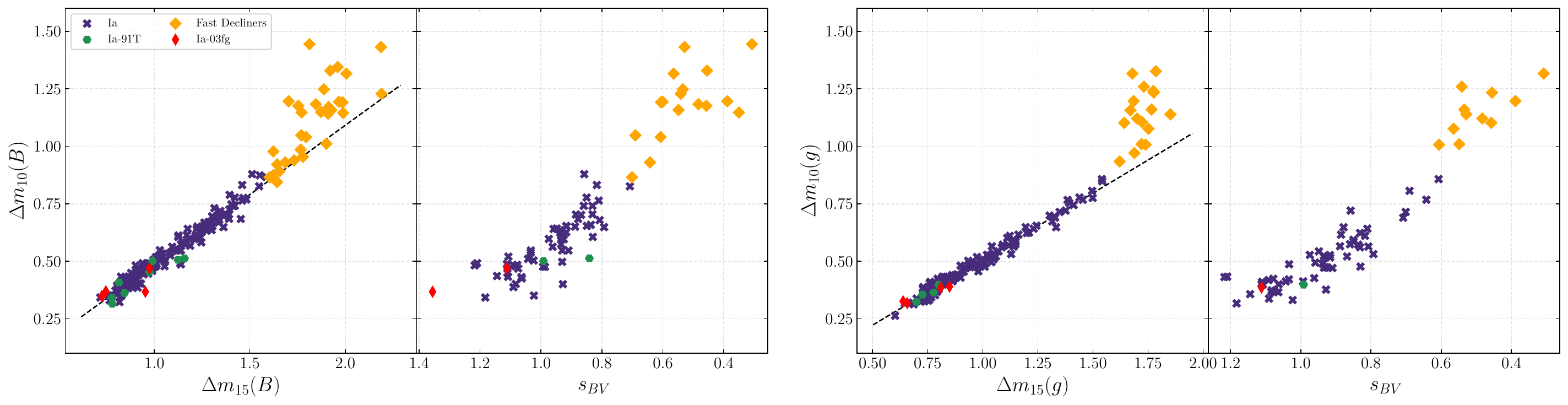} 
\caption{Direct comparisons of decline-rate and colour-stretch parameters. The panels are arranged horizontally, with the left two panels displaying $B$-band comparisons and the right two displaying $g$-band comparisons. From left to right, the panels show: $\Delta m_{10}(B)$ vs $\Delta m_{15}(B)$; $\Delta m_{10}(B)$ vs $s_{BV}$; $\Delta m_{10}(g)$ vs $\Delta m_{15}(g)$; and $\Delta m_{10}(g)$ vs $s_{BV}$. A linear relation (dashed line) is fitted to the normal SNe Ia ($\Delta m_{15}(B) < 1.6$\,mag) in the $\Delta m_{15}(X)$ panels to highlight the tight correlation for normal decliners, which breaks down for the fast-declining tail (orange diamonds). A legend detailing the subclass markers is provided in the far-left panel.}
\label{fig:dm15 vs dm10}
\end{figure*}

Another, highly relevant factor concerns the practicality of the approach; simply put, it is easier to get SN data over shorter lengths of time. Compared with $\Delta$$m$\textsubscript{$15$}($B$), for instance, less time coverage is required to compute a $\Delta$$m$\textsubscript{$10$}($B$) value. This means there is less of an opportunity for observing conditions to change which may impact measurements. It also allows for the possibility that previous SNe which lacked sufficient data for $\Delta$$m$\textsubscript{$15$}($B$) can be retrospectively studied should they meet the threshold of $\Delta$$m$\textsubscript{$10$}($B$). While selecting a value of $t$ lower than 10 would even further reduce required data, selecting too short a time range may also lead to unwanted effects. Definition of $t_{X,\text{max}}$ is crucial for ensuring an accurate measurement of light curve decline, and smaller values of $t$ would amplify these potential errors as a lesser value of $t$ is comparatively more difficult to interpolate. The error in defining $t_{X,\text{max}}$ will be constant, irrespective of the choice of $t$, hence the respective errors increasing.

\section{Discussion}
\label{sec:Discussion} 

Focusing on the chosen value of $t=10$, a direct comparison of the $\Delta m_{15}(B)$ and $\Delta m_{10}(B)$ LWR plots (Fig.~\ref{fig:phil grid}) demonstrates the clear advantage of the shorter timescale. It is worth reiterating the stark difference in fast-decliner behaviour between the two metrics: where the $\Delta m_{15}(B)$ relation exhibits a degenerate vertical break, $\Delta m_{10}(B)$ remains continuous and tracks the full luminosity range. While a transition `knot' is present, marking the shift to a steeper slope for fast decliners, the severe degeneracy demonstrated with $\Delta m_{15}(B)$ is resolved. Consequently, $\Delta m_{10}(B)$ acts as a superior tool for investigating transitional and fast-declining SNe at the faint end of the luminosity distribution.

This advantage is made clear by plotting the two decline-rates directly against one another (Fig.~\ref{fig:dm15 vs dm10}). For normal SNe Ia ($\Delta m_{15}(B) < 1.6$\,mag), there is a tight linear correlation between $\Delta m_{10}(B)$ and $\Delta m_{15}(B)$, confirming that these parameters can be safely converted into one another for slow and normal decliners. These conversions are defined as follows:
\begin{align}
\Delta m_{10}(B) &= 0.602 \cdot \Delta m_{15}(B) - 0.113 \label{eq:conv_B} \\
\Delta m_{10}(g) &= 0.574 \cdot \Delta m_{15}(g) - 0.065 \label{eq:conv_g}
\end{align}
However, as SNe cross the empirical fast-declining threshold, they deviate from this linearity. Because $\Delta m_{10}(B)$ maintains a continuous correlation with absolute magnitude, the clear break in linearity, seen in Fig.~\ref{fig:dm15 vs dm10}, confirms that $\Delta m_{15}(B)$ loses discerning power for fast-declining events. As shown in the bottom panels of Fig.~\ref{fig:dm15 vs dm10}, these exact trends are identically mirrored in the $g$-band.

\begin{figure*}
\centering
\includegraphics[width=0.86\textwidth]{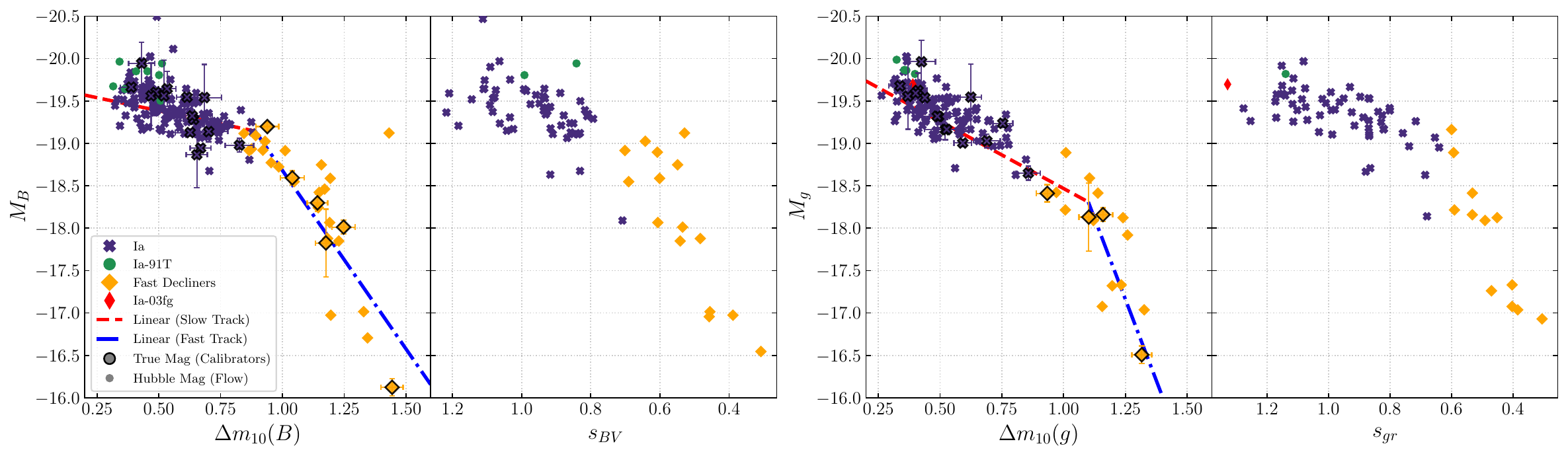}
\caption{Calibrated absolute magnitudes plotted against $\Delta m_{10}$ and colour-stretch parameters. The left two panels display $M_B$ versus $\Delta m_{10}(B)$ and $s_{BV}$, respectively. The right two panels display $M_g$ versus $\Delta m_{10}(g)$ and $s_{gr}$, respectively. True empirical calibrator SNe are plotted with bold black marker outlines and error bars. As in Fig.~\ref{fig:phil grid}, the broader `flow' sample is plotted as a faded background distribution utilising independent Hubble distances, with their error bars omitted for clarity. The derived continuous piecewise linear models (red for the normal track, blue for the fast track) are shown on the $\Delta m_{10}$ panels.}
\label{fig: combined}
\end{figure*}

To standardise SN Ia luminosities using $\Delta m_{10}(X)$, we fit the LWR using continuous piecewise linear models (plotted in Fig.~\ref{fig: combined}). The transition knots where the normal track meets the fast-declining track were empirically determined by our calibrator sample to be $\Delta m_{10}(B) = 0.89$\,mag and $\Delta m_{10}(g) = 1.10$\,mag.

For the $B$-band, the absolute magnitude is given by:
\begin{equation}
M_B = 
\begin{cases} 
-19.322 + 0.617\,(\Delta m_{10}(B) - 0.60) & \text{if } \Delta m_{10}(B) < 0.89 \\
-19.143 + 4.208\,(\Delta m_{10}(B) - 0.89) & \text{if } \Delta m_{10}(B) \ge 0.89
\end{cases}
\label{eq:wlr_B}
\end{equation}

For the $g$-band, the absolute magnitude is given by:
\begin{equation}
M_g = 
\begin{cases} 
-19.101 + 1.585\,(\Delta m_{10}(g) - 0.60) & \text{if } \Delta m_{10}(g) < 1.10 \\
-18.309 + 7.649\,(\Delta m_{10}(g) - 1.10) & \text{if } \Delta m_{10}(g) \ge 1.10
\end{cases}
\label{eq:wlr_g}
\end{equation}

The intrinsic root-mean-square (RMS) scatter around the normal, slow-declining track of our piecewise model is remarkably low: $0.08$\,mag in the $B$ band and $0.12$\,mag in the $g$ band. This confirms that $\Delta m_{10}(X)$ is a highly precise proxy for standardising normal SNe Ia. For comparison, fitting this exact same calibrator sample using the traditional $\Delta m_{15}(X)$ parameter yields a larger normal-regime scatter of $0.19$\,mag in the $B$ band and $0.30$\,mag in the $g$ band.

For the fast-declining tail ($\Delta m_{15}(B) \ge 1.6$\,mag), the intrinsic dispersion naturally increases, rising to $0.34$\,mag in the $B$ band and $0.40$\,mag in the $g$ band. However, this is a vast improvement over $\Delta m_{15}(X)$, which completely degenerates in this regime, yielding a RMS scatter of $0.94$\,mag and $0.75$\,mag in the $B$ and $g$ bands, respectively. While the piecewise model prevents these faint objects from breaking the overall calibration, this scatter is noticeably larger than the $0.19$\,mag RMS error achieved by \citet{phillips2025} when $s_{BV}$ is used as the light curve width indicator for fast decliners. This indicates that while $\Delta m_{10}$ is an excellent, short-baseline standardisation tool for normal SNe Ia, multi-band colour-stretch parameters like $s_{BV}$ remain superior for precisely resolving the luminosities of the fastest-declining events.

It is interesting to compare $\Delta m_{10}(X)$ and the colour-stretch parameters, $s_{BV}$ and $s_{gr}$ \citep{Burns_2014, Ashall2020}. Colour-stretch parameters are highly capable of tracking SN Ia brightness across the full magnitude range by utilising multi-band colour evolution. Shown in Fig.~\ref{fig: combined} is a comparison of $\Delta m_{10}(B)$ and $s_{BV}$ when plotted against $M_B$. While the correlations displayed are not identical, $\Delta m_{10}(B)$ offers a close approximation to the results of $s_{BV}$, despite its minimal data requirements. Fig.~\ref{fig:dm15 vs dm10} shows that $\Delta$$m$\textsubscript{10}($B$) smoothens the delineation caused by $\Delta$$m$\textsubscript{15}($B$) with $s$\textsubscript{$BV$} (Velasco et al., in prep). Furthermore, the relationship between $\Delta m_{10}(g)$ and $s_{gr}$ (Fig.~\ref{fig:dm15 vs dm10}) appears even tighter than its $B$-band counterpart across the entire luminosity range, suggesting that for slow declining SNe 
$\Delta m_{10}(B)$ may be advantageous when compared to $s_{BV}$.

This comparative behaviour extends to the $g$-band. As shown in Fig.~\ref{fig: combined}, $\Delta m_{10}(g)$ also provides a strong approximation of its corresponding colour-stretch parameter, $s_{gr}$. While colour-stretch parameters definitively perform better when tracking the luminosities of the fastest-declining SNe Ia at the faint end of the distribution, $\Delta m_{10}(X)$ provides a complementary alternative for the slow-decliners. When employed as a fast, low-effort tool, $\Delta m_{10}(X)$ has the capacity to replicate the standardisation power of a much more complex parameter.

As a final caveat, it should be noted that while our piecewise model successfully maps normal and fast-declining SNe, the peculiar SNe Ia-03fg are known outliers to many standardisation techniques \citep{Andrew_Howell_2006, ashall21}. For instance, while their slow decline-rates result in high predicted luminosities under our $\Delta m_{10}$ framework, objects such as ASASSN-15hy are physically underluminous \citep{Lu21}. As such they remain highlighted as a distinct subclass in our analysis.

\section{Conclusions}

We presented an explorative study of SNe Ia decline-rates utilising the $\Delta m_t(X)$ metric to evaluate alternative parameterisations of light curve diversity and the Luminosity-Width Relation. The primary advantage of $\Delta m_{10}(X)$ is not computational speed, but rather that it provides a robust, empirical indicator of light curve width across the full range of decline-rates without resorting to template fitting. By requiring only 10 days of post-maximum photometric coverage, $\Delta m_{10}(X)$ can be a highly practical tool for large-scale photometric surveys. It allows for rapid characterisation and absolute magnitude estimation with minimal data, making it a practical complement to more data-intensive metrics.

Our analysis confirms that while the traditional $\Delta m_{15}(B)$ parameter suffers from a degeneracy at the faint end of the luminosity distribution, $\Delta m_{10}(X)$ successfully mitigates this break, maintaining a continuous correlation with luminosity for both the $B$ and $g$ bands. 

However, we find that the intrinsic dispersion of the LWR is still dominated by the fast-declining tail ($\Delta m_{15}(B) \ge 1.6$\,mag). Because these rapidly evolving events (a regime encompassing both 1986G-like and 1991bg-like SNe) synthesise less $^{56}\mathrm{Ni}$ and exhibit fundamentally cooler, distinct photometric evolution \citep{10.1093/mnras/284.1.151}, they cannot simply be modelled as a linear extension of normal SNe Ia. Instead, they must be treated with a distinct calibration track. The chosen piecewise model successfully handles this transition without breaking the calibration, aligning with recent literature that also separately treats fast-decliners \citep{phillips2025, 2024MNRAS.530.4950G}.

Although $\Delta m_{10}(B)$ and $s_{BV}$ exhibit a tight relationship (Velasco et al., in prep), they possess distinct observational strengths. While multi-band colour-stretch parameters like $s_{BV}$ remain optimal for luminosity estimates of the faintest, fastest-declining SNe, $\Delta m_{10}(X)$ performs just as well, if not better, for the slow-declining extreme. With intrinsic dispersions of just $\sigma = 0.08$\,mag ($B$-band) and $0.12$\,mag ($g$-band) for normal SNe Ia, $\Delta m_{10}(X)$ provides a highly precise, short-baseline alternative for standardising the vast majority of the SN Ia population.

\section*{Acknowledgements}
The work of the CSP-II has been generously supported by the National Science Foundation under grants AST-1008343, AST-1613426, AST-1613455, and AST-1613472. The CSP-II was also supported in part by the Danish Agency for Science and Technology and Innovation through a Sapere Aude Level 2 grant. L. G. acknowledges financial support from CSIC, MCIN and AEI 10.13039/501100011033 under projects PID2023-151307NB-I00, PIE 20215AT016, CEX2020-001058-M, and by the MaX-CSIC Excellence Award MaX4-SOMMA-ICE. M. D. S. acknowledges support from the Independent Research Fund Denmark (IRFD; grant number 10.46540/2032-00022B). This research was supported by the Science and Technology Facilities Council (STFC) grants, whom J. M. V. would like to thank.

\textit{Facilities:} \textit{Magellan}:Baade (IMACS imaging spectrograph, FourStar wide-field near-infrared camera, FIRE near-infrared echellette), \textit{Magellan}:Clay (LDSS3 imaging spectrograph), Swope (SITe3 CCD imager, e2v 4 K $\times$ 4 K CCD imager), du Pont (Tek5 CCD imager, WFCCD imaging spectrograph, RetroCam near-infrared imager), Gemini:North (GNIRS near- infrared spectrograph), Gemini:South (FLAMINGOS2), VLT (ISAAC, MUSE), IRTF (SpeX near-infrared spectrograph), NOT (ALFOSC), Calar Alto 3.5 m (PMAS/PPak), La Silla-QUEST, CRTS, PTF, iPTF, OGLE, ASAS-SN, PS1, KISS, ISSP, MASTER, SMT). 

\textit{Software:} \textsc{SNID} \citep{2007ApJ...666.1024B}, \textsc{SUPERFIT} \citep{2005ApJ...634.1190H}, \textsc{GELATO} \citep{2008A&A...488..383HHaru}, \textsc{SNooPy} \citep{Burns2011AJ....141...19B, Burns_2014}.

\section*{Data Availability}

The CSP-I DR3 data used for this work is available at the Carnegie Supernova Project's respective website. CSP-II data is being released alongside relevant papers, though the whole survey in totality is still private. The datasets specifically curated for the analysis section, can be shared on reasonable request to the corresponding author. 



\bibliographystyle{mnras}
\bibliography{example} 





\bsp	
\label{lastpage}
\end{document}